 \documentclass[final,3p,times]{elsarticle}

 \usepackage{graphicx}

\usepackage{amssymb}
\usepackage{amsmath}

\usepackage{color}

\usepackage[dvips]{epsfig}
\usepackage{subfig}
\usepackage{float}
\usepackage{sidecap}

\usepackage{upgreek}
\usepackage{epstopdf}
\newcommand{\X}{\raisebox{2pt}{$\chi$}}
\usepackage{url}

\journal{Physica B: Condensed Matter}

\begin{document}

\begin{frontmatter}



\author[1]{Markus Gusenbauer\corref{email}}
\author[2,3]{Ha Nguyen}


\author[1]{Franz Reichel}
\author[1]{Lukas Exl}
\author[1]{Simon Bance}
\author[1]{Johann Fischbacher}
\author[1]{Harald \"Ozelt}
\author[1]{Alexander Kovacs}

\author[2]{Martin Brandl}
\author[1]{Thomas Schrefl}

\address[1]{Industrial Simulations, University of Applied Sciences, St. Poelten, Austria}
\address[2]{Center for Biomedical Technology, Danube University, Krems, Austria}
\address[3]{Institute for Microelectronics and Microsensors, Johannes Kepler University, Linz, Austria}
\cortext[email]{Corresponding author: \texttt{markus.gusenbauer@fhstp.ac.at}}


\title{Guided self-assembly of magnetic beads for biomedical applications}


\author{}

\address{}

\begin{abstract}

Micromagnetic beads are widely used in biomedical applications for cell separation, drug delivery, and hypothermia cancer treatment. Here we propose to use self-organized magnetic bead structures which accumulate on fixed magnetic seeding points to isolate circulating tumor cells. The analysis of circulating tumor cells is an emerging tool for cancer biology research and clinical cancer management including the detection, diagnosis and monitoring of cancer. Microfluidic chips for isolating circulating tumor cells use either affinity, size or density capturing methods. We combine multiphysics simulation techniques to understand the microscopic behavior of magnetic beads interacting with Nickel accumulation points used in lab-on-chip technologies. Our proposed chip technology offers the possibility to combine affinity and size capturing with special antibody-coated bead arrangements using a magnetic gradient field created by Neodymium Iron Boron permanent magnets. The multiscale simulation environment combines 
magnetic field computation, fluid dynamics and discrete particle dynamics. 

\end{abstract}


\end{frontmatter}


\section{Introduction}
\label{sec:intro}

The analysis of circulating tumor cells (CTCs) supports the monitoring of tumor growth and can be used to control the success of therapies. Microfluidic chips help to detect, to identify and to count these cells in peripheral blood. First time observed in 1869 \cite{ashworth_case_1869} it becomes possible to gain a better understanding of how metastases form through the analysis of circulating tumor cells with the advance of technology platforms. Due to their rare appearance existing microfluidic filters \cite{chen_microfluidic_2012} cannot find every single CTC in the blood flow. In these devises the distinct properties (size, affinity, density) of the tumor cells are used to filter them. The technical challenge is to detect, count and isolate one CTC over one billion cells \cite{yu_circulating_2011} (1--100 tumor cells per $ml$ blood).

A promising approach from Saliba et al. uses self organizing chains of ferromagnetic biofunctionalized beads \cite{saliba_microfluidic_2010}. An array of magnetic traps are prepared by microcontact printing in a microfluidic channel. Single particle chains line up which create a sieve like structure. This method has a limitation of flow rate due to decreasing stability with higher velocity. Our proposed chip technology uses very thin Nickel seeding points with a diameter several times larger then the bead diameter. In this work we will analyze the microfluidic behavior of softmagnetic beads attracted by this seeding points in multiphysics simulations. 


\begin{figure}[H]
\begin{center}
 \includegraphics[width=.48\textwidth]{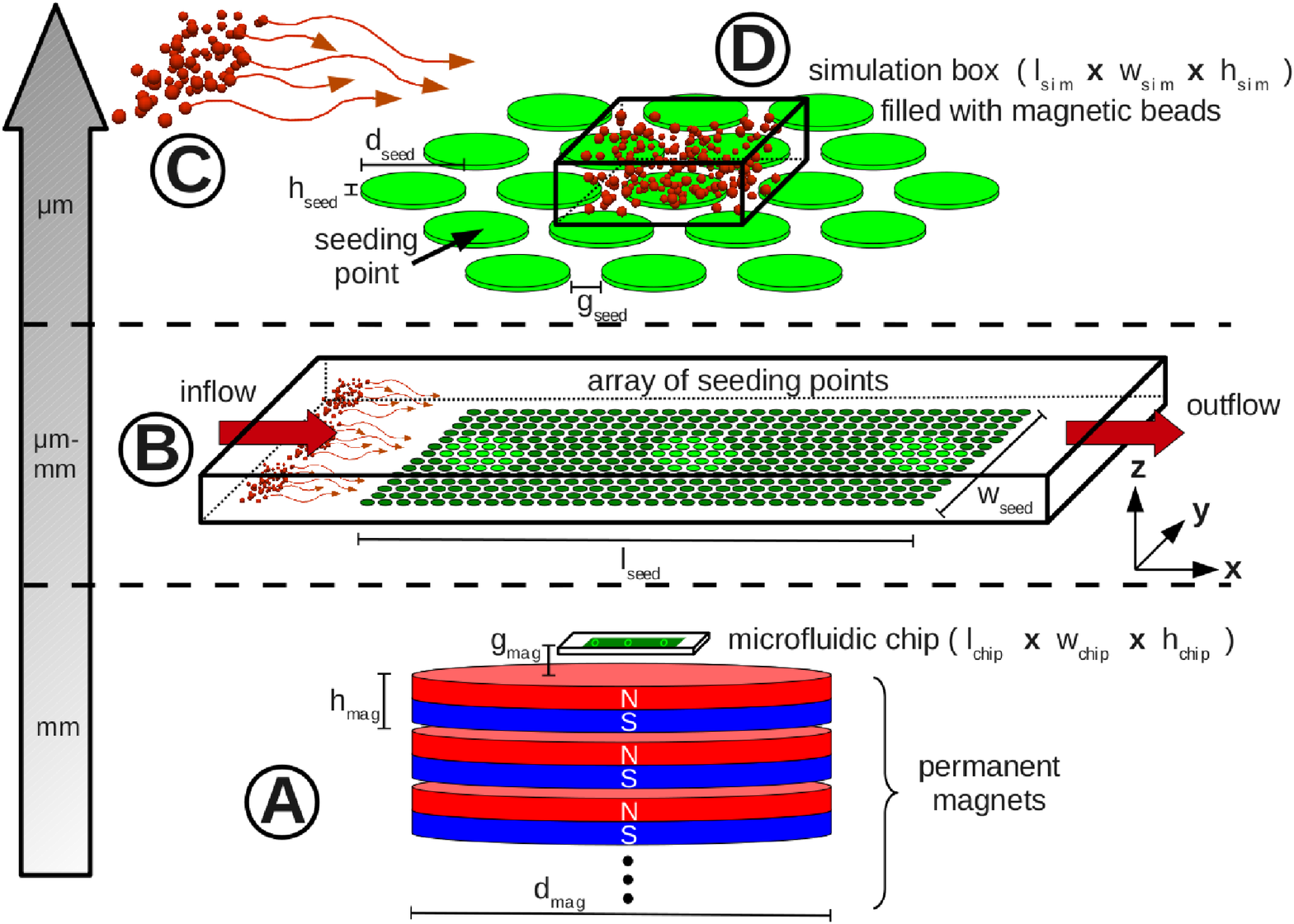}
\end{center}
 \caption{\small Multiscale simulation environment: \newline A) Magnetic field source: $N$ permanent magnets  B) Zoom into the microfluidic chip: Hexagonal array of Ni seedings points (light green - area of interest) C) Trajectories of softmagnetic particles D) Areas of interest: Magnetic field from permanent magnets and Ni discs for a closed simulation box filled with softmagnetic beads.}
 \label{fig:environment}
\end{figure}

\subsection*{Multiscale simulation environment}
\label{subsec:content}
A microfluidic chip (Fig. \ref{fig:environment}A) is placed on top of a single or multiple permanent magnets. On the bottom of the chip a hexagonal array of Nickel cylinders is placed. The cylinders act as a accumulation point (seeding point) for softmagnetic particles in the fluid flow. When applying the permanent magnets on the microfluidic chip the softmagnetic particles self-organize according to the field created by the seeding points and the permanent magnets. The behavior of magnetic beads close to a single seeding point is discussed in Section \ref{subsec:beadLocationSeed}.

Cuboidal or cylindrical NdFeB permanent magnets are the magnetic source for the given scenario. Akoun \cite{akoun_3d_1984} showed the analytic calculation of the magnetic field created by a cuboidal permanent magnet. Derby \cite{derby_cylindrical_2010} did the same for cylindrical permanent magnets. To get a higher magnetic field $\vec H$ several magnets are combined having superposition of the field values (Fig. \ref{fig:environment}A)


In order to reduce simulation time and computational cost we are focusing on special areas in the microfluidic channel. Fig. \ref{fig:environment}B shows the microfluidic chip with the seeding point array at the bottom. The light green seeding points will be taken into account for further investigation. The simulation boundaries are set according to the results of Section \ref{subsec:seedSeed}.

The magnetic field $\vec H$ magnetizes the interacting seeding points. They create a highly non-homogeneous magnetic field in the microfluidic chip. Fig. \ref{fig:environment}D shows the simulation area close to a single seeding point. Softmagnetic particles are randomly filled into the simulation box and interact with each other, the permanent magnets and the seeding points.

Looking at a larger scale the trajectories of the magnetic beads to the seeding points are calculated with the software package Comsol \cite{comsol} (Fig. \ref{fig:environment}C) . The particle distribution, i.e. the number of beads close to a single seeding point, depends on the fluid velocity, the applied magnetic field, the chip geometry as well as the seeding material.

\begin{table}[h]
\small
\begin{centering}
\begin{tabular}{lrl}
  \textbf{magnetic particles} & &\\
  diameter $d_{\mathrm{bead}}$ & 10 & $\upmu$m\\
  mass magnetization $\sigma$ & 25--35 & A m$^\textrm{2}$ kg$^\textrm{-1}$\\
  \noalign{\smallskip}\hline\noalign{\smallskip}
  \textbf{cyl. permanent magnets} & &\\
  diameter $d_{\mathrm{mag}}$ & 60 & mm\\
  height $h_{\mathrm{mag}}$ & 5 & mm \\
  gap to chip $g_{\mathrm{mag}}$ & 0.5--100 & mm \\
  magnetic polarization $J_{\mathrm{s}}$ & 1.6 & T\\
  \noalign{\smallskip}\hline\noalign{\smallskip}
  \textbf{Ni seeding points} & &\\
  diameter seeding point $d_{\mathrm{seed}}$ & 300 & $\upmu$m\\
  height seeding point $h_{\mathrm{seed}}$ & 1--100 & $\upmu$m\\
  gap between seeding points $g_{\mathrm{seed}}$ & 50 & $\upmu$m\\
  unaxial anisotropy constant $K1$ & 5.7 & kJ m$^\textrm{-3}$\\
  magnetic polarization $J_\mathrm{s}$ & 0.62 & T\\
  exchange constant $A$ & 90 & pJ m$^\textrm{-1}$\\
  Gilbert damping constant $\alpha $ & 0.1 & \\
  \noalign{\smallskip}\hline\noalign{\smallskip}
  \textbf{simulation box} & &\\
  geometry $l_{\mathrm{sim}} \times w_{\mathrm{sim}} \times h_{\mathrm{sim}}$ & 500$\ \times\ $500$\ \times\ $100 & $\upmu$m\\
  grid length $l_{\mathrm{g,sim}}$ & 5 & $\upmu$m\\
  \noalign{\smallskip}\hline
\end{tabular}
 \caption{\small Parameters of the simulation environment}
 \label{tab:param}
 \end{centering}
\end{table}

\section{Methods}
\label{sec:methods}

\subsection{Magnetic particle dynamics}
\label{subsec:particleDynamics}

Under the influence of a magnetic field a magnetic moment $m$ is created in every particle. With the moments of two nearby beads and the distance $r$ we got a formulation (Eqn. \ref{eqn:interaction}) of the interaction force $F_\mathrm{i}$ for bead 2 and vice versa for bead 1 \cite{furlani_permanent_2001}. 
\begin{align}
\begin{split}
\vec F_{1\rightarrow 2}= \frac{3\mu_0}{4\pi r^{5}}\bigg[(\vec m_1 \cdot \vec r)\vec m_2 + (\vec m_2 \cdot \vec r)\vec m_1 + (\vec m_1 \cdot \vec m_2)\vec r- \frac{5(\vec m_1 \cdot \vec r)(\vec m_2 \cdot \vec r)}{r^{2}}\vec r\bigg]
\end{split}
\label{eqn:interaction}
\end{align}

The gradient force $\vec F_\mathrm{g}$ (Eqn. \ref{eqn:Fstart}) on a bead is given by the negative gradient of the energy of the magnetic dipole moment $\vec m$ in the field $\vec B$. Eqn. \ref{eqn:Fstart} shows the 3-dimensional vector with the assumption of homogeneous magnetization inside the beads ($\partial_{\vec r}\ \vec m=0$).
\begin{align}
\vec F_\mathrm{g}=\nabla(\vec m \cdot \vec B)=\begin{pmatrix} 
						m_x\partial_xB_x+m_y\partial_xB_y+m_z\partial_xB_z \\
						m_x\partial_yB_x+m_y\partial_yB_y+m_z\partial_yB_z \\
						m_x\partial_zB_x+m_y\partial_zB_y+m_z\partial_zB_z
					      \end{pmatrix}
\label{eqn:Fstart}
\end{align}



In our preliminary work we assumed that the external field is inhomogeneous only in one dimension and also the field itself has only a single direction \cite{gusenbauer_selforganizing_2012}. This assumption reduces computational time a lot. We derived only the z-field in y-direction leaving a resulting force $F_{\mathrm{g},z}=m_z\partial_yB_z$. But if we want to have a complex magnetic field from several permanent magnets and thin Ni discs, we have to consider all 3 dimensions.

During simulation this force could be calculated in every timestep for every single bead which slows down the simulation. Another possibility is the initial calculation of the external field $\vec H$ and its derivatives $\partial_iB_j$ at the beginning of the simulation in a Cartesian grid. And during the simulation this fixed values are interpolated according to the real bead position in the grid.


We implemented both methods, the direct calculation in every timestep, and the particle-in-cell method for faster computation in the open-source particle simulator YADE \cite{kozicki_new_2008}.

\subsection{Particle-in-cell method}
\label{subsec:particleInCell}
\begin{figure}[H]
\centering
  \subfloat[]{\epsfig{file=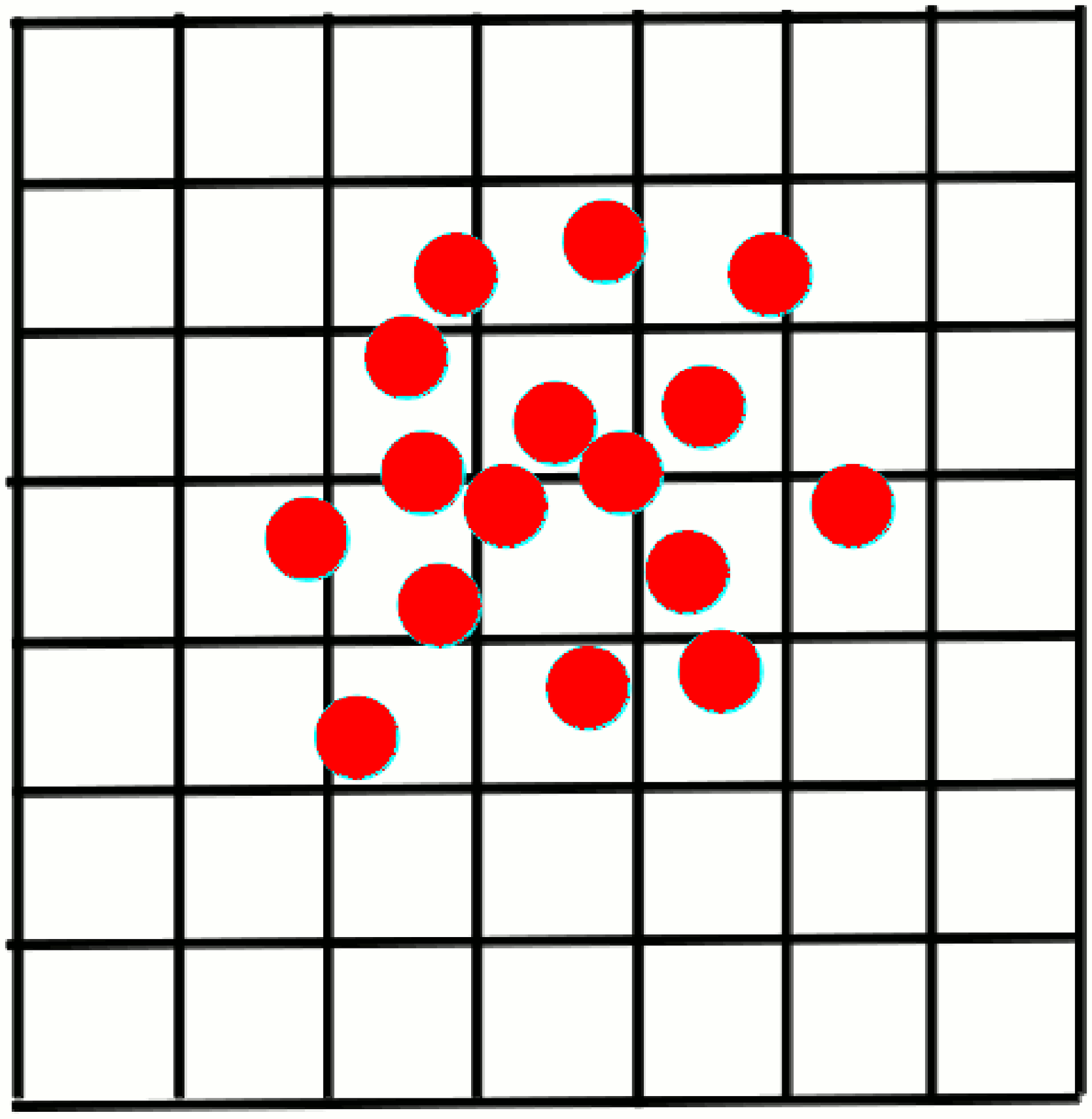,width=0.1\textwidth}\label{subfig:grid}}\qquad
  \subfloat[]{\epsfig{file=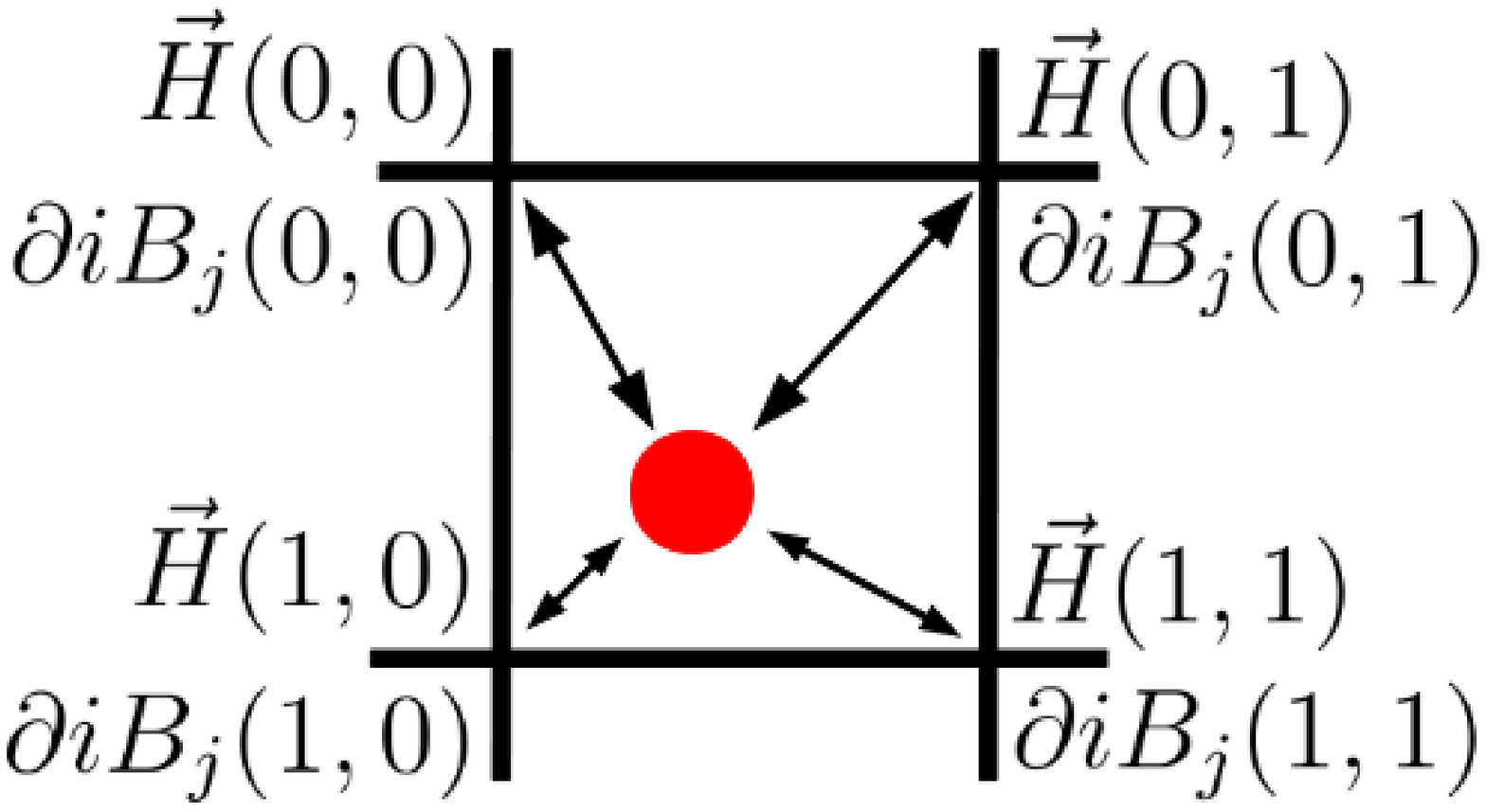,width=0.2\textwidth}\label{subfig:interpolate}}\qquad
\caption{
\small   \protect\subref{subfig:grid} Fixed Cartesian grid with magnetic particles. \protect\subref{subfig:interpolate} Interpolation of field and derivatives according to particle position. }
\label{fig:partInCell}
\end{figure}

The particle-in-cell method works with a fixed Cartesian grid (Fig. \ref{subfig:grid}). In our case the simulation box in Fig. \ref{fig:environment}D is the boundary of the particle simulation and contains the fixed grid with uniform grid length $l_{\mathrm{g,sim}}$. Before the actual simulation of interacting magnetic beads, initial values for the external field $\vec H$ and all derivatives $\partial_iB_j$ need to be calculated in every single grid node. To get the magnetic field values in the grid several steps are performed.

\begin{enumerate}
 \item Analytic calculation of the magnetic field from permanent magnets at seeding points of interest using \cite{akoun_3d_1984} for cuboidal and \cite{derby_cylindrical_2010} for cylindrical permanent magnets. 
 \item Numerical calculation of the field $\vec H$ in every grid point of the simulation box using the finite element micromagnetic package FEMME \cite{femme}. The amount of seeding points taken into account for the field calculation in the simulation box are derived in Section \ref{subsec:seedSeed}.
 \item Numerical differentiation of the grid to get all values of $\partial_iB_j$ using three point formulas \cite{economics_numericaldiff}. Eqn. \ref{eqn:3pointcommon} shows a common three point formulation that includes the point before $(x_i-l_\mathrm{g,sim})$ and after $(x_i+l_\mathrm{g,sim})$ the calculated point $x$ (centered difference formula). At the edge points Eqn. \ref{eqn:3pointedge} is used as there are no points outside the grid (forward or backward difference formula). Eqn. \ref{eqn:3pointcommon} and \ref{eqn:3pointedge} generate a second-order accurate approximation with a truncation error of $O(l^2_\mathrm{g,sim})$.
\end{enumerate}
\begin{align}
\partial_iB_j(x_i) \approx \frac{B_j(x_i+l_{\mathrm{g,sim}}) - B_j(x_i-l_{\mathrm{g,sim}})}{2l_{\mathrm{g,sim}}} \label{eqn:3pointcommon} \\
\partial_iB_j(x_i) \approx \frac{-3 B_j(x_i) + 4 B_j(x_i\pm l_{\mathrm{g,sim}}) - B_j(x_i\pm 2 l_{\mathrm{g,sim}})}{2 l_{\mathrm{g,sim}}} \label{eqn:3pointedge}
\end{align}
During the simulation in every timestep the following tasks are performed to get the magnetic induced force on the beads

\begin{enumerate}
 \item Interpolate magnetic field $\vec H$ and derivatives $\partial_iB_j$ for every single magnetic bead according to position in grid cell using tricubic interpolation (Eqn. \ref{eqn:interpolate} and Fig. \ref{subfig:interpolate}).
 \begin{align}
f(x,y,z)=\sum\limits_{i,j,k=0}^3 a_{ijk} x^i y^j z^k
\label{eqn:interpolate}
\end{align}
 \item Calculate magnetic moment in every single bead ($\vec M=\X \vec H$) using interpolated field $\vec H$ from the grid
 \item Calculate and add force on bead due to magnetic particle interaction (Eqn. \ref{eqn:interaction})
 \item Calculate and add force on bead due to magnetic gradient field (Eqn. \ref{eqn:Fstart}) using interpolated derivatives $\partial_iB_j$ from the grid
\end{enumerate}


The simulation takes into account the influence from the permanent magnets and the seeding points onto the magnetic grid at the beginning of the simulation. There is no change of magnetic grid values from magnetic beads or seeding points during the simulation run. Eqn. \ref{eqn:interaction} handles bead-bead interactions, therefore the grid node values stay the same. Regarding the change of magnetic field originated by the interaction of magnetic beads and the seeding points Section \ref{subsec:beadSeed} gives the answer. There is only little change of magnetic field in the seeding points caused by the interacting magnetic beads, hence the resulting magnetic field can be neglected.

\subsection{Particle trajectories}
\label{subsec:partTraject}
This work focuses on the magnetic particle behavior close to the seeding points. In order to understand the particle flow towards the point of interest (Fig. \ref{fig:environment}C)  the simulation environment could be enlarged, which would need enormous computational power. Another possibility was proposed from members of our project consortium \cite{nguyen_dynamic_2012}. They calculate the particle trajectory with the Lagrange equation in COMSOL multiphysic \cite{comsol}. 2D simulations and more realistic 3D simulations are possible. 
\begin{align}
\rho _\mathrm{p} V_\mathrm{p} \frac{d\vec u_\mathrm{p}}{dt}=C_\mathrm{d} \frac{\rho _\mathrm{f} A_\mathrm{p}}{2} |\vec u_\mathrm{f} - \vec u_\mathrm{p}| (\vec u_\mathrm{f} - \vec u_\mathrm{p}) + \vec F
\label{eqn:Lagrange}
\end{align}

The Lagrange equation (Eqn. \ref{eqn:Lagrange}) incorporates the velocity from the particle relative to the fluid $\vec u_\mathrm{f} - \vec u_\mathrm{p}$, all other forces $\vec F$ coming basically from magnetic gradient fields and particle properties (density $\rho _\mathrm{p}$, volume $V_\mathrm{p}$). Solving this equation allows to predict the particle movement inside the microfluidic device and the distribution along the seeding points.

\section{Results}
\label{sec:results}

\subsection{Seed-Seed-Interaction}
\label{subsec:seedSeed}

We will investigate the center of the array where the magnetic field from the permanent magnets points nearly only into z-direction (Fig. \ref{fig:environment}B). The light green areas at the inflow and the outflow are influenced by all three magnetic field components. $H_y$ and $H_z$ components are equal for inflow and outflow whereas $H_x$ values are mirrored at the z-axis. Therefore only the inflow or the outflow need to be considered in the simulation as one side is just the reflected copy of the other one.


The simulation box for the magnetic bead simulation (Fig. \ref{fig:environment}D) is placed on top of a seeding point in the area of interest. Neighboring seeding points interact with the center Ni disc. The magnetic field $\vec H$ at several positions inside the center seed is summed up from the analytically calculated field $\vec H_i$ of varying neighboring seeds. They are assumed to have maximum magnetic polarization $J=J_\mathrm{s}=\ $0.62 T. With increasing distance to the center seed, i.e. after the second circle of seeding points, the influence gets insignificant. Therefore the field calculation for the simulation box done with FEMME is limited to 18 surrounding seeding points

\subsection{Bead-Seed-Interaction}
\label{subsec:beadSeed}

The particle-in-cell method requires field and derivatives calculation at the beginning of the simulation in every grid point of the simulation box. During simulation the grid values could be changed due to magnetized objects (beads, seeding points) inside the simulation environment. The interaction of magnetic beads is already handled with Eqn. \ref{eqn:interaction}. Seeding points could create a significant change of the magnetic field. Without magnetic beads the field created by the seeding points is caused only by the applied permanent magnets and the interaction with other seeding points. 


Interacting magnetic beads create again a magnetic field that changes magnetization inside the seeding points. The maximum field created by magnetic beads is insignificant (100 times smaller) compared to the field created by the permanent magnets. The theoretical amount of magnetic beads in the simulation box is more than 40 000 beads, which we will never get, because we want to create a filter structure for circulating tumor cells. 

\subsection{Magnetic beads close to seeding point}
\label{subsec:beadLocationSeed}

\begin{figure}[h]
\centering
  \subfloat[]{\epsfig{file=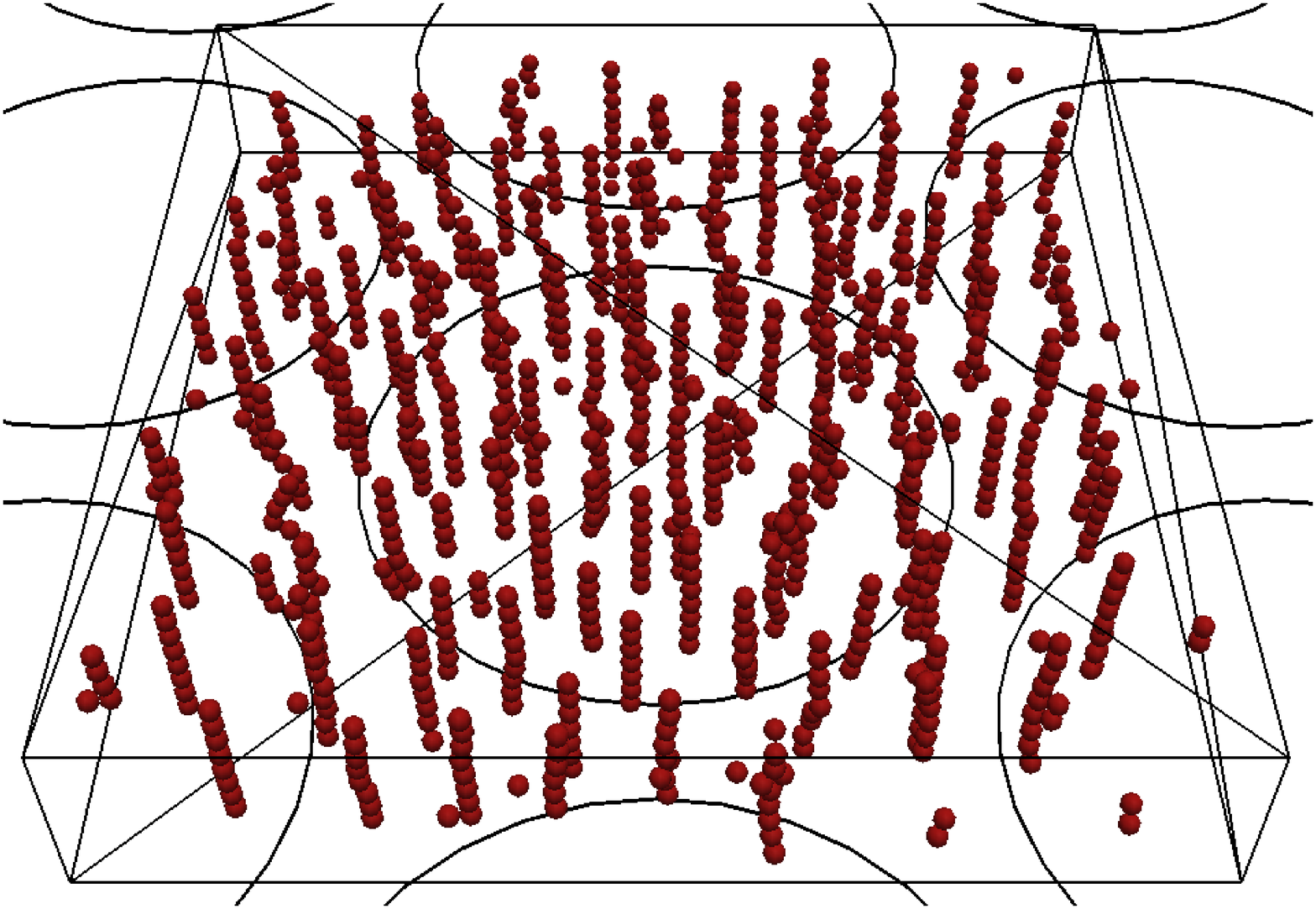,width=0.2\textwidth}\label{subfig:compPermClose}}\qquad
  \subfloat[]{\epsfig{file=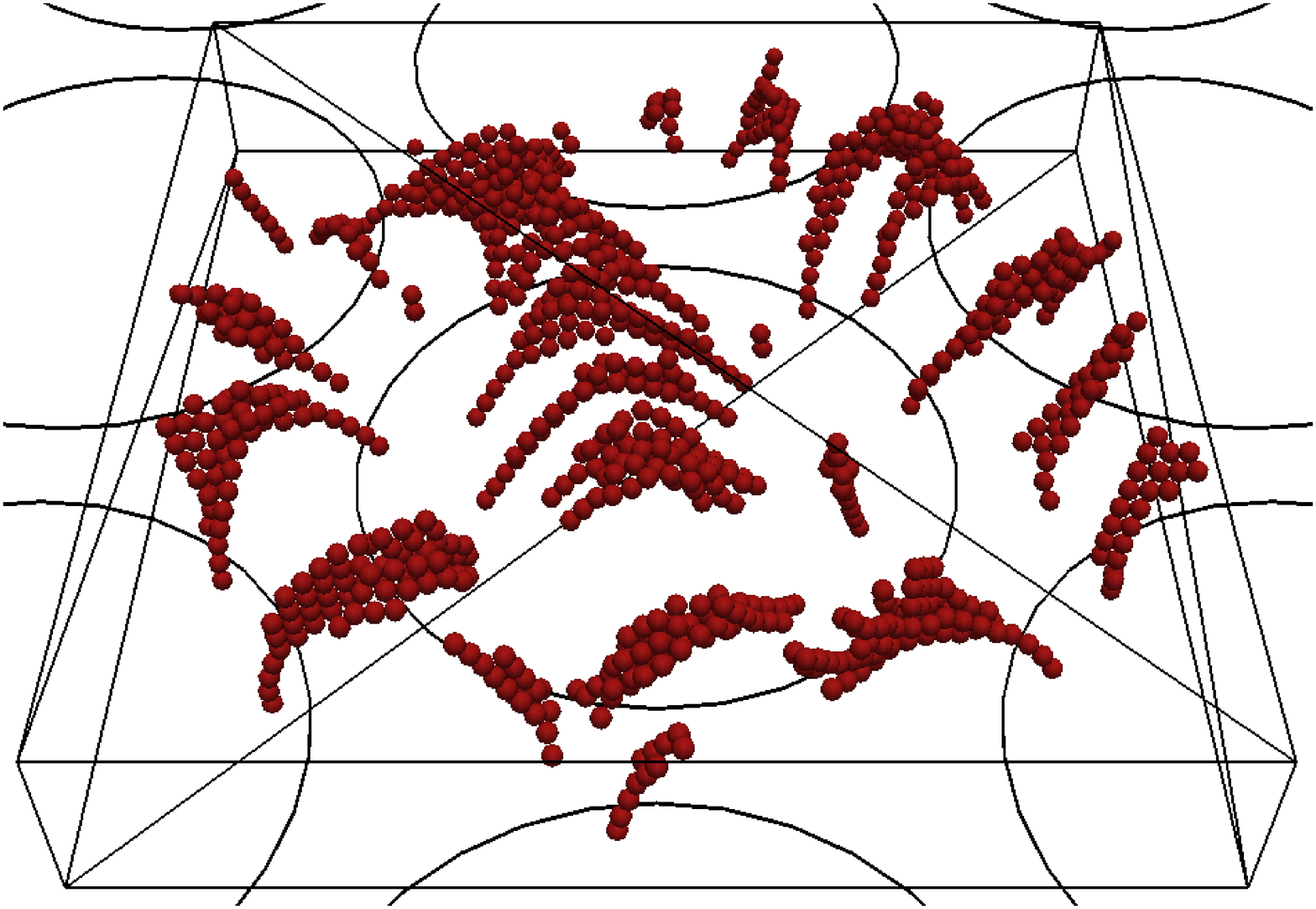,width=0.2\textwidth}\label{subfig:compHomogen}}\qquad
\caption{
\small Simulation box filled with softmagnetic beads. Resulting bead structures from 1 permanent magnet in 
0.5 mm distance to the microfluidic chip  \protect\subref{subfig:compPermClose} and from a homogeneous magnetic field ($B_z=\ $0.45 T) \protect\subref{subfig:compHomogen} with subjacent Ni seeding array.}
\label{fig:comp}
\end{figure}

Simulations with 1 permanent magnet (parameters from Table \ref{tab:param}) with a certain distance to the chip ($g_{\mathrm{mag}}=\ $10 cm) does not essentially move the magnetic beads. Only little bead agglomerations occur without a significant influence of the Ni seeding points. Moving the permanent magnet closer to the chip increases the field at the seeding array. Consequently the field gradient increases too, let the beads start to interact more. The strength of the permanent magnet field gradient compared to the gradient created by the magnetized seeding points prevents the interaction with the seeds. Magnetic bead chains are created independent of the subjacent Ni array ($g_{\mathrm{mag}}=\ $ 0.5 mm, Fig. \ref{subfig:compPermClose}). In between $g_{\mathrm{mag}}=\ $10 cm and $g_{\mathrm{mag}}=\ $0.5 mm the beads undergo a smooth transition from little agglomerations to chains, never influenced by the seeding points. 

In order to have filter structures specially tuned with the Ni array the strong magnetic field gradient of the permanent magnet needs to be erased. Permanent magnets with a lower magnetic polarization $J_\mathrm{s}$ do not change this fact. Applying a homogeneous magnetic field ($\nabla \vec B = 0$) fulfills the requirement for tuning the filter device (e.g. from a horseshoe magnet or Halbach-cylinder \cite{halbach_design_1980}). Fig. \ref{subfig:compHomogen} shows the strong interaction from the magnetic beads close to the seeding points. They align according to the field lines from the array. A similar result is expected replacing the Ni seeds using micro-patterning of permanent magnets \cite{dempsey_hard_2009} without applying an external field. 

\section{Conclusion}
\label{sec:conclusion}
Circulating tumor cells are captured due to mechanical or affinity properties. In this work we try to combine both methods with softmagnetic bead arrays. Other than in our preliminary work \cite{gusenbauer_selforganizing_2012}, where magnetic beads self-organize just by a magnetic gradient field and the fluid flow, the arrangement is guided with a regular pattern of thin Ni discs. The array is magnetized with applied NdFeB permanent magnets or homogeneous magnetic fields. To understand the behavior of the beads in the microfluidic device a simulation environment combining macro- and microscale was developed. Analytical field calculations of the applied permanent magnets are the base for numerical analysis of the seeding array. The importance of field strength is shown on the dominant magnetization of the thin Ni discs as well as the neglectable field from bead agglomerations acting on the seeding points.

Magnetic particle dynamics with a large number of beads can get computational expensive. It can be controlled by limiting interaction radii of every single bead. Calculation of magnetic field in every single bead is also time consuming. Therefore a particle-in-cell method was implemented into the simulation environment. Thus the field calculations from permanent magnets and Ni seeding points do not need to be calculated in every timestep. During initialization of the simulation a regular grid is filled with magnetic field values. Additional numerical derivatives complete the Cartesian grid. During simulation magnetic field values and derivatives for every single bead are gained through tricubic interpolation.

Resulting softmagnetic bead structures have shown to be independent on the Ni array using permanent magnets as external field source. High magnetic field gradients suppress relatively weak seeding field gradients. Ni patterns affect the resulting bead structure much more using a homogeneous field source. The investigation of the microfluidic device can be extended using different materials and geometries.

Which particular setup is most suitable for filtering CTCs need to be examined in further studies. Blood flow simulations (see preliminary work for details \cite{gusenbauer_selforganizing_2012, cimrak_modelling_2012, gusenbauer_simulation_2013}) help to improve the design of the device for optimal bead arrangements. \\

\small{\textbf{Acknowledgment} The authors gratefully acknowledge the financial support of the N\"O Forschungs- und Bildungsges.m.b.H. (NFB) through the Life Science Calls}




\bibliographystyle{model1a-num-names}
\bibliography{hmm2013_gusenbauer}

\begin{thebibliography}{17}
\expandafter\ifx\csname natexlab\endcsname\relax\def\natexlab#1{#1}\fi
\providecommand{\bibinfo}[2]{#2}
\ifx\xfnm\relax \def\xfnm[#1]{\unskip,\space#1}\fi
\bibitem[{Ashworth(1869)}]{ashworth_case_1869}
\bibinfo{author}{T.~R. Ashworth}, \bibinfo{journal}{Aust Med J}
  \bibinfo{volume}{14} (\bibinfo{year}{1869}) \bibinfo{pages}{146–149}.
\bibitem[{Chen et~al.(2012)Chen, Li, and Sun}]{chen_microfluidic_2012}
\bibinfo{author}{J.~Chen}, \bibinfo{author}{J.~Li}, \bibinfo{author}{Y.~Sun},
  \bibinfo{journal}{Lab on a Chip} \bibinfo{volume}{12} (\bibinfo{year}{2012})
  \bibinfo{pages}{1753--1767}.
\bibitem[{Yu et~al.(2011)Yu, Stott, and Toner~et al.}]{yu_circulating_2011}
\bibinfo{author}{M.~Yu}, \bibinfo{author}{S.~Stott},
  \bibinfo{author}{M.~Toner~et al.}, \bibinfo{journal}{The Journal of Cell
  Biology} \bibinfo{volume}{192} (\bibinfo{year}{2011})
  \bibinfo{pages}{373--382}.
\bibitem[{Saliba et~al.(2010)Saliba, Saias, and Psychari~et
  al.}]{saliba_microfluidic_2010}
\bibinfo{author}{A.-E. Saliba}, \bibinfo{author}{L.~Saias},
  \bibinfo{author}{E.~Psychari~et al.}, \bibinfo{journal}{Proceedings of the
  National Academy of Sciences} \bibinfo{volume}{107} (\bibinfo{year}{2010})
  \bibinfo{pages}{14524--14529}.
\bibitem[{Akoun and Yonnet(1984)}]{akoun_3d_1984}
\bibinfo{author}{G.~Akoun}, \bibinfo{author}{J.~Yonnet},
  \bibinfo{journal}{{IEEE} Transactions on Magnetics} \bibinfo{volume}{20}
  (\bibinfo{year}{1984}) \bibinfo{pages}{1962--1964}.
\bibitem[{Derby and Olbert(2010)}]{derby_cylindrical_2010}
\bibinfo{author}{N.~Derby}, \bibinfo{author}{S.~Olbert},
  \bibinfo{journal}{American Journal of Physics} \bibinfo{volume}{78}
  (\bibinfo{year}{2010}) \bibinfo{pages}{229}.
\bibitem[{com(2013)}]{comsol}
\bibinfo{title}{{COMSOL} 4.2 release highlights},
  \bibinfo{howpublished}{\url{http://www.comsol.com/products/4.2/}},
  \bibinfo{year}{2013}. \bibinfo{note}{Last visited on 08/05/2013}.
\bibitem[{Furlani(2001)}]{furlani_permanent_2001}
\bibinfo{author}{E.~P. Furlani}, \bibinfo{title}{Permanent magnet and
  electromechanical devices: materials, analysis, and applications},
  \bibinfo{publisher}{Academic Press}, \bibinfo{year}{2001}.
\bibitem[{Gusenbauer et~al.(2012)Gusenbauer, Kovacs, and Reichel~et
  al.}]{gusenbauer_selforganizing_2012}
\bibinfo{author}{M.~Gusenbauer}, \bibinfo{author}{A.~Kovacs},
  \bibinfo{author}{F.~Reichel~et al.}, \bibinfo{journal}{Journal of Magnetism
  and Magnetic Materials} \bibinfo{volume}{324} (\bibinfo{year}{2012})
  \bibinfo{pages}{977--982}.
\bibitem[{Kozicki and Donz{\'e}(2008)}]{kozicki_new_2008}
\bibinfo{author}{J.~Kozicki}, \bibinfo{author}{F.~Donz{\'e}},
  \bibinfo{journal}{Computer Methods in Applied Mechanics and Engineering}
  \bibinfo{volume}{197} (\bibinfo{year}{2008}) \bibinfo{pages}{4429--4443}.
\bibitem[{fem(2013)}]{femme}
\bibinfo{title}{{FEMME} software  - suessco.com},
  \bibinfo{howpublished}{\url{http://suessco.com/simulations/solutions/femme-s%
oftware/}}, \bibinfo{year}{2013}. \bibinfo{note}{Last visited on 08/05/2013}.
\bibitem[{eco(2013)}]{economics_numericaldiff}
\bibinfo{title}{Economics 613/614 computational methods for economics and
  applications fall 2007, notes on numerical differentiation},
  \bibinfo{howpublished}{\url{http://www.karenkopecky.net/Teaching/eco613614/N%
otes_NumericalDifferentiation.pdf}}, \bibinfo{year}{2013}. \bibinfo{note}{Last
  visited on 02/05/2013}.
\bibitem[{Nguyen et~al.(2012)Nguyen, Rauch, and Brandl}]{nguyen_dynamic_2012}
\bibinfo{author}{H.~Nguyen}, \bibinfo{author}{C.~Rauch},
  \bibinfo{author}{M.~Brandl}, \bibinfo{publisher}{{ACTAPRESS}},
  \bibinfo{year}{2012}, pp. \bibinfo{pages}{764--154:1}.
\bibitem[{Halbach(1980)}]{halbach_design_1980}
\bibinfo{author}{K.~Halbach}, \bibinfo{journal}{Nuclear Instruments and
  Methods} \bibinfo{volume}{169} (\bibinfo{year}{1980}) \bibinfo{pages}{1--10}.
\bibitem[{Dempsey(2009)}]{dempsey_hard_2009}
\bibinfo{author}{N.~M. Dempsey}, in: \bibinfo{booktitle}{Nanoscale Magnetic
  Materials and Applications}, \bibinfo{publisher}{Springer {US}},
  \bibinfo{year}{2009}, pp. \bibinfo{pages}{661--683}.
\bibitem[{{Cimrak} et~al.(2012){Cimrak}, Gusenbauer, and
  Schrefl}]{cimrak_modelling_2012}
\bibinfo{author}{I.~{Cimrak}}, \bibinfo{author}{M.~Gusenbauer},
  \bibinfo{author}{T.~Schrefl}, \bibinfo{journal}{Comput. Math. Appl.}
  \bibinfo{volume}{64} (\bibinfo{year}{2012}) \bibinfo{pages}{278–288}.
\bibitem[{Gusenbauer et~al.(2013)Gusenbauer, \"Ozelt, and Fischbacher~et
  al.}]{gusenbauer_simulation_2013}
\bibinfo{author}{M.~Gusenbauer}, \bibinfo{author}{H.~\"Ozelt},
  \bibinfo{author}{J.~Fischbacher~et al.}, \bibinfo{journal}{{EPJ} Web of
  Conferences} \bibinfo{volume}{40} (\bibinfo{year}{2013})
  \bibinfo{pages}{02001}.

\end{thebibliography}







\end{document}